\documentclass[letterpaper, english, aps, pra, twocolumn, floatfix, showpacs, amsfonts, amssymb, superscriptaddress]{revtex4-1}
\usepackage{epsfig, graphicx,float}
\usepackage[T1]{fontenc}
\usepackage[latin1]{inputenc}
\usepackage{bm}
\usepackage{color}
\usepackage{babel}
\usepackage{hyperref}
\usepackage{psfrag}
\usepackage{amsmath}
\usepackage{amssymb}

\begin{document}

\title{Phase transition of ultracold atoms immersed in a BEC vortex lattice}
\author{R. H. Chaviguri}
\affiliation{Instituto de F\'{i}sica de S\~ao Carlos, Universidade de S\~ao Paulo, C.P. 369, S\~ao Carlos, SP, 13560-970, Brazil}
\author{T. Comparin}
\affiliation{Laboratoire de Physique Statistique, \'Ecole Normale Sup\'erieure/PSL Research University,
UPMC, Universit\'e Paris Diderot, CNRS, 24 rue Lhomond, 75005 Paris, France}
\author{V. S. Bagnato}
\affiliation{Instituto de F\'{i}sica de S\~ao Carlos, Universidade de S\~ao Paulo, C.P. 369, S\~ao Carlos, SP, 13560-970, Brazil}
\author{M. A. Caracanhas}
\affiliation{Instituto de F\'{i}sica de S\~ao Carlos, Universidade de S\~ao Paulo, C.P. 369, S\~ao Carlos, SP, 13560-970, Brazil}

\begin{abstract}
We investigate the quantum phases of ultracold atoms trapped in a vortex lattice
using a mixture of two bosonic species (A and B), in the presence of an
artificial gauge field. Heavy atoms of species B are confined in the array of
vortices generated in species A, and they are described through a Bose-Hubbard
model.  In contrast to the optical-lattice setups, the vortex lattice has an
intrinsic dynamics, given by its Tkachenko modes. Including these quantum
fluctuations in the effective model for B atoms yields an extended Bose-Hubbard
model, with an additional ``phonon''-mediated long-range attraction.  The
ground-state phase diagram of this model is computed through a variational
ansatz and the quantum Monte Carlo technique. When compared with the ordinary
Bose-Hubbard case, the long-range interatomic attraction causes a shift and
resizing of the Mott-insulator regions.  Finally, we discuss the experimental
feasibility of the proposed scheme, which relies on the proper choice of the
atomic species and on a large control of physical parameters, like the
scattering lengths and the vorticity.
\end{abstract}

\date{\today}
\maketitle

\section{Introduction}

Two-component condensate systems had an enormous impact in the field of
ultracold atoms, especially since the experimental realization of a
Bose-Einstein condensate (BEC) of fermions. This consists of effective bosonic
molecules, the tightly bound Cooper pairs, formed after the fermions be
sympathetically cooled by another bosonic or fermionic atomic species.  The
experimental observation of the BEC-BCS crossover with this molecular BEC
\cite{zwerger1} proved once more the relevance of ultracold atoms as a powerful
tool to test condensed-matter models, which can be studied in a highly
controllable environment \cite{ibloch, ibloch2003}.  The two-species BEC of
bosonic atoms also has rich physics to be explored. An important experiment
performed in Cornell's group, with different $^{87}$Rb hyperfine states,
addressed the static properties of binary mixtures, their relative phase
coherence and their dynamics \cite{cornell1}. The same group was able to
nucleate vortices in this system \cite{cornell2}, and, more recently, to produce
the superposition of vortex-lattice BECs \cite{cornell3}.

The vortex configuration in BEC was predicted by Feynman \cite{feynman}, who
suggested that a superfluid can rotate when pierced by an array of quantized
singularities or vortices. In 1969, Tkachenko proposed that a vortex lattice in
a superfluid would support transverse elastic-modes \cite{tkachenko}. He showed
that a triangular lattice has the lowest energy of all simple lattices with one
vortex per unit cell, and that it is stable for all normal modes.  These
predictions were experimentally realized with dilute BEC gases in 2001
\cite{ketterle}, and followed by the observation of vortex-line oscillations,
that is, the low-energy Tkachenko modes \cite{cornell4}.  Once established, the
vortex-lattice configuration proved to be stable, including its normal modes,
for a range of rotation frequencies close to, but below that of the external
trap confining potential \cite{cornell5,fetter1}.

In a previous work \cite{pereira}, we considered a neutral impurity immersed in
a vortex lattice, interacting with the Tkachenko modes. We addressed the
shallow-lattice regime, assuming a quadratic energy dispersion for the impurity
species, and taking the continuous limit for its momentum.  In this work, an
analogous system is studied in the tight-binding regime, where heavy atoms
(impurities) are strongly trapped in the sites of the vortex lattice formed by
condensed bosons of another species.  This vortex-lattice setup should be
compared to the technique of trapping ultracold atoms in optical lattices
generated by laser beams. The study of bosonic atoms in optical lattices led to
the breakthrough observation of the quantum phase transition between the
superfluid and Mott-insulating phases \cite{greiner}, well described by the
Bose-Hubbard (BH) model for lattice bosons \cite{zoller3}.

Important results obtained with optical lattices rely on the fact that these are
rigid (i.e., do not support phonons \cite{Lewenstein2012}) and free of defects.
To better simulate the condensed-matter models for solid-state crystals,
however, a recent trend has been to introduce artificial dynamics in these light
crystals \cite{Griessner2007NJP, Klein2007PRA, Bruderer2008PRA}.  In our
proposed scheme, the dynamics emerges naturally through the normal-modes
excitations of the vortex lattice. The effective Hamiltonian for the
lattice-confined impurities corresponds to a polaronic Bose-Hubbard (BH) model,
with parameters that are modified by the lattice dynamics, notably by the
introduction of a long-range ``phonon''-mediated interaction.

The generalization of BH models in the ultracold atoms context is connected to
the use of atoms with strong dipole moment, for which the strength of
interatomic interaction decays slowly with the distance \cite{Goral2002PRL}. In
this case, the system is described through an extended Bose-Hubbard (EBH) model,
with the additional long-range interaction generating a rich variety of phases.
For the one-dimensional case, the phase diagram includes the peculiar
Haldane-insulator phase \cite{DallaTorre2006PRL}. In two dimensions, the
predicted phases range from density-wave to supersolidity \cite{Lahaye2009RPP,
Trefzger2011JPB}.  Ref.~\cite{ferlaino}, in particular, reports the experimental
realization of a EBH model in three dimensions.

The vortex-lattice setup proposed here also realizes a EBH models, due to the
effective long-range attraction generated by the lattice dynamics.  Similarly to
optical lattices, the high and independent control of the system parameters
would allow one to explore the quantum phases of impurities in a vortex lattice.
The relevant experimental methods include the use of Feshbach resonances to tune
atomic scattering lengths \cite{Chin2010RMP} and of the selective
absorption-image technique to characterize the quantum-state configuration
\cite{esslinger, jin2008, ibloch2003}.

This paper is structured as follows: Sec.~\ref{sect:model} presents the
derivation of an effective Bose-Hubbard model for impurity atoms trapped by the
vortex lattice of the other atomic species. The phase diagram of this model is
determined through the quantum Monte Carlo technique, and compared to the
existing results (\emph{cf.} Sec.~\ref{sect:BH}).  In Sec.~\ref{sect:dynamics},
we give a beyond mean-field treatment for species A, which leads to the EBH
model for impurity atoms, including long-range interactions.  The phase diagram
of this ``dynamical'' model is analyzed in Sec.~\ref{sect:EBH}. In
Sec.~\ref{sect:experiments}, we describe the physical parameters relevant for an
experimental realization of our proposal, and the main conclusions are reported
in \ref{sect:conclusions}.

\section{Physical model}
\label{sect:model}

We consider a two-component mixture of species A and B, in a quasi-2D geometry.
The vortex lattice in species A is excited by the artificial-gauge-field
technique \cite{Dalibard2011RMP}, where the internal atomic structure is
carefully engineered by the optical potentials to produce Berry phases and
nucleate vortices \cite{spielman}. The main idea is to replace the rotating trap
mechanism by using an artificial vector potential $\mathbf{A}$, which
selectively couples to A atoms and does not affect species B.  The Hamiltonian
${H}$ of the system is the sum of the following three terms:
\begin{eqnarray}
\label{eq1}
\nonumber
{H}_A &=&
\int d^2r\,\Big[
\hat\psi_A^\dagger(\mathbf{r})
\frac{(-i\hbar\nabla-{\bf A(\mathbf{r})})^2}{2 m_A}
\hat\psi_A^{\phantom\dagger}(\mathbf{r}) + \\
\nonumber
&&
+ \hat\psi_A^\dagger(\mathbf{r}) V_{\textrm{ext}}(\mathbf{r}) \hat\psi_A(\mathbf{r})
+ \frac{g_{A}}{2} \left(\hat\psi_A^{\dagger}(\mathbf{r})\hat\psi_A(\mathbf{r})\right)^2 \Big], \\
\nonumber
{H}_B &=&
\int d^2r\,\Big[
\hat\psi_B^\dagger(\mathbf{r})
\frac{(-i\hbar\nabla)^2}{2 m_B}
\hat\psi_B(\mathbf{r}) + \\
\nonumber
&&
+ \hat\psi_B^\dagger(\mathbf{r}) V_{\textrm{ext}}(\mathbf{r}) \hat\psi_B(\mathbf{r})
+ \frac{g_B}{2} \left(\hat\psi_B^{\dagger}(\mathbf{r}) \hat\psi_B(\mathbf{r})\right)^2 \Big], \\
{H}_{AB} &=& g_{AB} \int d^2r\,
\hat\psi_A^{\dagger}(\mathbf{r}) \hat\psi_B^{\dagger}(\mathbf{r})
\hat\psi_A (\mathbf{r}) \hat\psi_B(\mathbf{r}),
\label{eq:va1}
\end{eqnarray}
where species $i \in \lbrace A, B\rbrace$ is described by the creation
(destruction) field-operator $\hat{\psi}^{\dagger}_{i}(\mathbf{r}) \;
(\hat{\psi}_{i}(\mathbf{r}))$ at the two-dimensional position $\mathbf{r} =
(x,y)$.  The strength of intra and inter-species repulsive contact interactions
are given by $g_{i}={2\sqrt{2\pi} \hbar^2\,a_{i}}/{m_{i} \ell^i_{z}}$ and
$g_{AB}={\sqrt{2\pi}\hbar^2\,a_{AB}}/{m_{AB} \ell^{AB}_{z}}$, respectively,
where $m_{AB}={m_A m_B}/{(m_A + m_B)}$ is the reduced mass, and $a_i$ and
$a_{AB}$ are the intra and inter-species s-wave scattering lengths.  The length
scales $\ell^{i}_{z} = \sqrt{{\hbar}/(m_{i}\omega_{z})}$ and $\ell^{AB}_{z} =
\sqrt{{\hbar}/{(m_{AB}\omega_{z})}}$ are functions of the harmonic confining
frequency $\omega_z$ in the transverse direction. The vector potential
$\mathbf{A}$ determines the vorticity $\mathbf{\Omega}$ for the BEC A along the
$z$ axis through $\Omega=|\nabla\times\mathbf{A}|/m_A$. At a critical vorticity,
the residual confining potential for A tends to zero \cite{matveenko2009}. For
species B, the effect of the slowly-varying external trap potential
$V_\mathrm{ext}(\mathbf{r})$ may be included through the Local Density
Approximation.

Within a mean-field treatment of ${H}_A$, the vortex-lattice wave function
for species A is built as the linear combination of degenerate
lowest-Landau-levels solutions of the rotational Gross-Pitaevskii equation.
This yields an Abrikosov lattice of vortices, encoded in the wave function
$\psi_{A}(\mathbf{r}) = \sqrt{n_A} \varphi_A(\mathbf{r})$, where $n_A = N_A /S$
is the average atomic density (with $S$ being the surface area). The
$\mathbf{r}$-dependent factor reads \cite{matveenko2009, matveenko2011}
\begin{equation}
\label{eq:vortex_mf}
\varphi_A(\mathbf{r}) =
(2\upsilon)^{1/4}\,
\vartheta_{1}(\zeta \sqrt{\pi\upsilon}, \rho)\,
\exp{\frac{\zeta^2-|\zeta|^{2}}{2}},
\end{equation}
where $\zeta$ represents the complex variable $(x + \imath y) / l_z^A$, $\rho =
\exp(\imath \pi
\tau)$, $\tau = u + \imath \upsilon$, $u = -1/2$ and $\upsilon = \sqrt{3} / 2$.
The triangular lattice of vortices is formed by the zeros of the Jacobi
theta function $\vartheta_{1}$.

For the stability of this vortex lattice we assume a vanishing temperature,
$T=0$. Then, in spite of not having a phase-coherent system, as part of the
atoms A are outside the condensate state, we still have well established
vortex-lattice density profile \cite{sinova}.  Moreover, we consider the
mean-field quantum-Hall regime \cite{matveenko2011}, with the number of vortices
$N_V$ well below the number of atoms in A, that is, a high ratio $\nu = N_A/N_V
\gg 1$.  In this regime, neither quantum nor thermal fluctuations affect the
vortex-lattice stability. We also assume $N_B \sim N_V \ll N_A$, that allow us
to disregard the effects of the dilute species on the stability of the vortex
lattice.

In the above scenario, we can apply the Bogoliubov transformation in the field
operator $\hat\psi_A$ to derive the excitations around the vortex-lattice
fundamental state, that is, to include the quantum fluctuations beyond the
mean-field approximation for A: $\hat{\psi}_{A} = \psi_{A} +
\delta\hat{\psi}_{A}$. We consider the grand-canonical formalism, with chemical
potentials $\mu_i$ for the species A and B, and rewrite the total system
Hamiltonian in Eq.~\eqref{eq:va1} as an expansion in powers of
$\delta\hat{\psi}_{A}$. To the second-order, this expansion reads
\begin{equation}
\label{eq:va2}
K =
\underbrace{H_{B}+H_{AB}^{(0)}-\mu_{B}\hat{N}_{B}}_{K_B} +
\underbrace{H_{AB}^{(1)}}_{H_\mathrm{int}} +
\underbrace{H_{A}^{(2)}-\mu_{A}\hat{N}_{A}}_{K_A^\mathrm{BOG}}
\end{equation}
where $\hat{N}_{A}$ and $\hat{N}_{B}$ are the number operators for species A and
B, and where the superscripts in $H_A^{(2)}$, $H_{AB}^{(0)}$ and $H_{AB}^{(1)}$
indicate the order of the expansion. Considering the validity of the mean-field
vortex-lattice solution for species A, the coefficient of the first order term
$H_A^{(1)}$ is zero. We truncate the interaction term $H_{AB}$ to first order,
and we do not show here the mean-field contribution to the energy of species A.

In this section we only consider the $0$th order term in Eq.~\eqref{eq:va2},
$K_B$, in which species A appears as an effective mean-field potential for
species B.  The inter-species interaction term reads $H_{AB}^{(0)}=\int d^2 r\,
V_A(\mathbf r) \hat{\psi}_B^\dagger (\mathbf r) \hat{\psi}_B^{\phantom\dagger}
(\mathbf r)$, where $V_A(\mathbf r) = n_A g_{AB} |\varphi_A(\mathbf r)|^2$ with
$\varphi_A(\mathbf r)$ given by Eq.~\eqref{eq:vortex_mf}, constitutes the static
lattice potential seen by B atoms.  In the dilute regime for species B (i.e.,
for $N_B \ll N_A$), the repulsive interspecies interaction causes the
localization of B atoms at the vortex-core positions, which is energetically
favorable as the density of A atoms vanish there.  This effect is at the the
basis of the derivation of an effective Bose-Hubbard model for B atoms, as
detailed below.

We focus on the tight-binding regime, where the lattice depth $V_0 = n_A g_{AB}$
is much larger than the recoil energy $E_r = \hbar^2/ (2 m_B \xi^2)$.  $E_r$ is
the natural energy-scale for B atoms trapped in a vortex core, with radius
approximately equal to the healing length $\xi$($= \hbar / \sqrt{2 m_A n_A
g_A})$ of the species A BEC.  The parameter $\Gamma_\mathrm{LLL}=n_A
g_{A}/2\hbar\Omega$ is associated to the lowest-Landau-level constraint
($\Gamma_\mathrm{LLL}<1$, \emph{cf.} Ref.~\cite{cornell5}), and it connects with
the vortex-lattice density $n_V$ through $n_V \sim 1 / \pi d^2$, where $d = 2
\sqrt{\hbar / (m_A \Omega)}$ is the inter-vortex separation
\cite{feynman,fetter1}.  We assume that the energy-level spacing between the
Bloch bands is large compared to the relevant energies of processes involving B
atoms, which are then restricted to the lowest-energy band. The single-band
assumption is especially valid in the tight-binding regime ($V_0 / E_r \gg 1$)
considered in this work. This allows us to expand the field operator for species
B in terms of Bloch wave functions:
\begin{equation}
\label{eq:va3}
\hat{\psi}_{B}(\mathbf{r}) =
\sum_{\mathbf{k}}\Phi_{\mathbf{k}}(\mathbf{r})\,\hat{b}_{\mathbf{k}},
\end{equation}
where $\hat{b}_{\mathbf{k}}$ destroys a particle in a quasi-momentum state
$\mathbf{k}$.  The Wannier function $\varphi_{B}(r_j)$, defined through
$\Phi_{\mathbf{k}}(\mathbf{r}) =({1}/{\sqrt{N_{V}}}) \sum_j \,\varphi_{B}(r_j)\,
e^{\imath \mathbf{k} \cdot \mathbf{R}_j}$, is localized at the vortex sites
$\mathbf{R}_{j} $ $(r_j = |\mathbf{r}-\mathbf{R}_j|)$ and is normalized to one.
Expanding the field operator as a sum of Wannier functions in each lattice site,
and considering only nearest-neighbor hopping and on-site interaction, $K_B$ can
be rewritten as a BH Hamiltonian \cite{zoller3}:
\begin{equation}
\label{eq:va4}
K_B =
- J \sum_{\langle i,j \rangle} \hat{b}_{i}^{\dag} \hat{b}_{j}
+ \frac{U}{2} \sum_{i} \hat{n}_i(\hat{n}_i-1)
-\mu_{B} \sum_{i} \hat{n}_i,
\end{equation}
where $\hat{b}_{i} = ({1}/{\sqrt{N_{V}}})\sum_{\mathbf{k}}
e^{\imath\mathbf{k}\cdot\mathbf{R}_{i}}\hat{b}_{\mathbf{k}}$, $\hat{n}_i =
\hat{b}_i^\dag \hat{b}_i$, and $\langle i,j \rangle$ denotes nearest-neighbor
pairs.
The hopping coefficient and on-site repulsion strength are given respectively by
\begin{equation}
J = -\int d^{2}r\,
\varphi^{\ast}_B(r_i) \left[
-\frac{\hbar^{2}\nabla^{2}}{2m_{B}} + g_{AB} n_{A} |\varphi_A(\mathbf{r})|^{2}
\right] \varphi_{B}(r_j),
\end{equation}
and
\begin{equation}
U = g_{B} \int d^{2}r \, |\varphi_B(r)|^{4}.
\label{eq:va5}
\end{equation}
$U$ can be determined by assuming a Gaussian function for $\varphi_{B}(r)\,(=
|B_{0}|\,e^{-r^2/2\ell_{0}^{2}}\,$) \cite{ibloch3}, with $\int\, d^2 r \,
|\varphi_{B} (r)|^2 = 1$ and width $\ell^{2}_{0}={\hbar\xi}/{\sqrt{m_B V_0}}$
(harmonic approach for the vortex-core density profile \cite{fetter2}). This
gives $U = {g_{B}}/{2 \pi\ell^{2}_{0}}$.  On the other hand, the Gaussian ansatz
leads to a poor approximation for $J$, which is better determined through the
solution of a 1D Mathieu equation \cite{zwerger2}.  In analogy with the case of
optical lattices, the resulting expression for $J$ reads
\begin{equation}
\label{eq:va6}
J = \frac{4}{\sqrt{\pi}} E_{r} \left(\frac{V_{0}}{E_{r}}\right)^{\frac{3}{4}}
\exp\left(-2\sqrt{\frac{V_{0}}{E_{r}}}\right).
\end{equation}
Using the vortex-lattice potential depth and recoil energy in the expressions
for $J$ and $U$, we obtain
\begin{equation}
\label{eq:va7}
\frac{U}{J} = \frac{1}{\sqrt{2}}
\frac{a_B}{\ell^B_z}
\left(\frac{a_A}{2a_{AB}}\frac{m_{AB}}{m_B}\right)^{1/4}
e^{\sqrt{\frac{2a_{AB}}{a_A} \frac{m_B}{m_{AB}}}}.
\end{equation}
In contrast with optical-lattice setups, where $U/J$ is controlled by the
laser-beam properties and atomic scattering length, in the vortex lattice this
ratio is connected to the atomic properties through the inter- and intra-species
scattering lengths ($a_A$, $a_B$, and $a_{AB}$), which can be independently
controlled with uniform magnetic fields via the Feshbach-resonance technique
\cite{Chin2010RMP}.  The mapping in Eq.~\eqref{eq:va7} opens the possibility of
using the vortex-lattice setup described in this paper to explore the BH phase
diagram on a triangular lattice, as will be shown in Sec.~\ref{sect:BH}.

\section{Bose-Hubbard phase diagram}
\label{sect:BH}

\begin{figure}[hbt]
\centering
\includegraphics[width=0.98\linewidth]{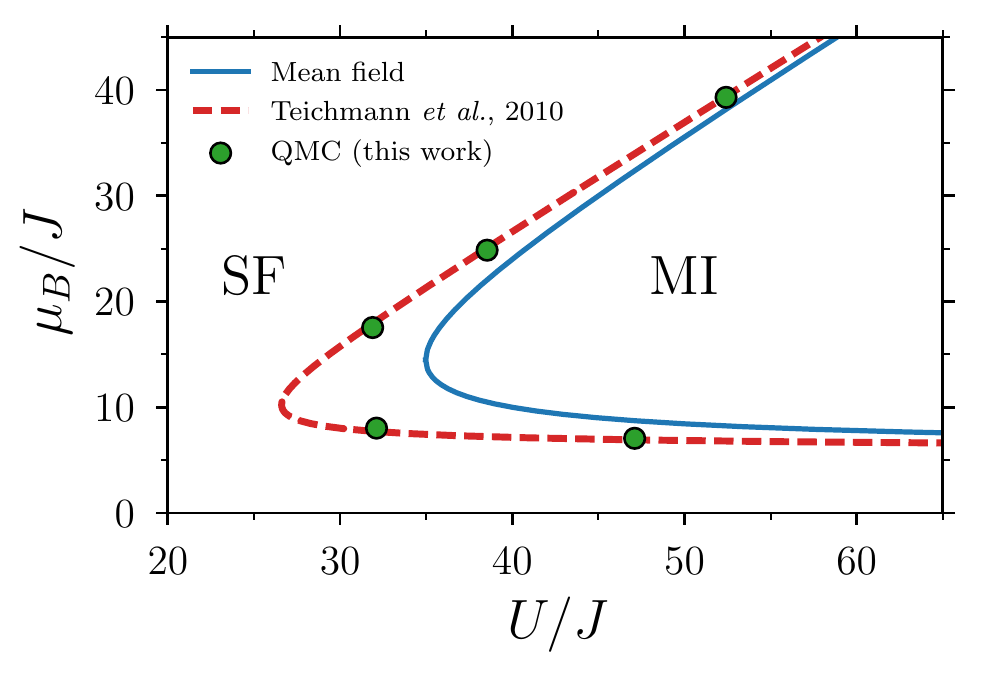}
\caption{Phase boundary between the Mott-insulator lobe with filling factor
$g=1$ (MI) and the superfluid phase (SF), on a two-dimensional triangular
lattice.  Different boundaries are obtained through mean-field theory (blue
solid line, see Eq.~\eqref{eq:BH_mf}), process-chain approach (dashed red line,
from Ref.~\cite{Teichmann2010EPL}), and the quantum Monte Carlo technique (green
circles, \emph{cf.} Appendix \ref{AppendixQMC}).
\label{fig:BH}}
\end{figure}

At zero temperature, the Bose-Hubbard model in Eq.~(\ref{eq:va4}) features a
phase transition between a Mott insulator (MI), with an integer number of atoms
per site, and a phase-coherent superfluid (SF) phase \cite{fisher}. The phase
diagram has been characterized in detail, for several geometries and
dimensionalities (see Ref.~\cite{Krutitsky2016PR} for a review).  For the
two-dimensional triangular lattice, one can compute the phase boundaries at
several levels of approximation.  The simplest case consists in a mean-field
approach \cite{zoller3}. By decoupling the kinetic part of Eq.~(\ref{eq:va4})
through $\hat{b}_{i}^{\dagger}\,\hat{b}_{j} \simeq \langle \hat{b}_{i}^{\dagger}
\rangle \hat{b}_{j} +\hat{b}_{i}^{\dagger}\,\langle \hat{b}_{j}\rangle - \langle
\hat{b}_{i}^{\dagger} \rangle\langle \hat{b}_{j} \rangle $, the Hamiltonian
becomes a sum of single-site terms: $K_B = \sum_i K_B^{(i)}$. Assuming a real
and homogeneous local order parameter $\Psi = \langle \hat{b}_{i} \rangle$, each
term reads
\begin{equation}
\label{eq:va8}
K_B^{(i)} =
- z J \Psi (\hat{b}_{i}^{\dagger} +\hat{b}_{i}) + z J \Psi^2 +
\frac{U}{2} \hat{n}_i(\hat{n}_i-1) - \mu_B \hat{n}_i,
\end{equation}
where $z$ is the number of neighbors per site ($z=6$, for a triangular lattice).
The first term in Eq.~\eqref{eq:va8} is then treated through second-order
perturbation theory \cite{fisher, stoof}. We expand the resulting energy
spectrum in powers of the order parameter $\Psi$, and apply Landau criterion to
identify the phase transition \cite{stoof}. The boundary between the superfluid
phase and the Mott insulator with filling factor $g$ is given by
\begin{equation}
\label{eq:BH_mf}
\mu_B =
\frac{U(2 g-1) - z J}{2} \pm
\frac{\sqrt{U^2- 2 z U J (2g + 1)+z^2 J^2}}{2},
\end{equation}
where the $+$ ($-$) sign refers to the upper (lower) boundary of the
Mott-insulator lobe (\emph{cf.} Fig.~\ref{fig:BH}).

The mean-field phase diagram is known to underestimate the area of the
Mott-insulator lobes.  More accurate boundaries are obtained through the
diagrammatic process-chain approach, where a perturbation series in $J$ is
computed up to high order.  On the square lattice, this technique yields a phase
diagram in extremely good agreement with unbiased quantum Monte Carlo (QMC)
results \cite{Teichmann2009PRB}.  Results of the process-chain approach are also
available on the triangular lattice \cite{Teichmann2010EPL}, for which the tip
of the Mott-insulator lobe with filling $g=1$ is at $(U/J,
\mu_B/J)_\mathrm{crit}\simeq(26.6, 10.2)$ -- \emph{cf.} Fig.~\ref{fig:BH}.

We also study the triangular-lattice BH model through the worm-algorithm QMC
technique \cite{Albuquerque2007JMMM, Bauer2011JStatMech}.  We extract the
critical points through the finite-size-scaling analysis of the superfluid
density (\emph{cf.} Appendix \ref{AppendixQMC}), and the numerical results are
in good agreement with the process-chain phase boundary (see Fig.~\ref{fig:BH}).
In Sec.~\ref{sect:EBH}, the same QMC algorithm is used to characterize the
extended Bose-Hubbard model.

\section{Vortex-lattice dynamics}
\label{sect:dynamics}

The model derived and characterized in Sections~\ref{sect:model} and
\ref{sect:BH} does not include a peculiar aspect of the vortex-lattice physics,
namely the intrinsic dynamics determined by its normal modes.  The inclusion of
these modes modifies the hopping amplitude and the interactions for B atoms.  In
addition to the trapping mechanism which keeps B atoms in the vortex lattice of
A, we also consider the scattering of the B atoms by the Tkachenko modes of
species A.  In the following, we derive the effective BH Hamiltonian that
follows from the inclusion of the lattice ``vibrations'', i.e., the quantum
fluctuations beyond the mean-field vortex-lattice solution in
Eq.~\eqref{eq:vortex_mf}.

According to Eq.~\eqref{eq:va2}, the first-order term in powers of
$\delta\hat{\psi}_{A}$ yields \cite{pereira}
\begin{equation}
\label{eq:va9}
H_\mathrm{int} =
g_{AB} \int d^{2}r
\left[
\delta\hat{\psi}_{A}^{\dag} \hat{\psi}_{B}^{\dag} \hat{\psi}_{B} \psi_{A} +
\psi^{\ast}_{A} \hat{\psi}_{B}^{\dag}\hat{\psi}_{B} \delta\hat{\psi}_{A}
\right].
\end{equation}
The second-order term, i.e., the Hamiltonian $K^\mathrm{BOG}_{A}$ for species A,
is diagonalized through the Bogoliubov-mode expansion
\begin{equation}
\delta\hat\psi_A(\mathbf{r}) =
\frac{1}{\sqrt{S}}
\sum_{\mathbf{q}} \left[
u_{\mathbf{q}}(\mathbf{r})\hat{a}_{\mathbf{q}} -
v_{\mathbf{q}}(\mathbf{r})\hat{a}_{\mathbf{q}}^{\dagger}
\right].
\end{equation}
This expansion leads to $K^\mathrm{BOG}_{A} = \sum_{\mathbf{q}} \epsilon_{q}
\hat{a}_{\mathbf{q}}^{\dagger}\hat{a}_{\mathbf{q}}$, plus a constant term that
only shifts the mean-field chemical potential.  This expression is obtained for
specific values of $u_{\mathbf{q}}$ and $v_{\mathbf{q}}$, and it includes the
operator $\hat{a}^{\dagger}_\mathbf{q}$ ($\hat{a}_\mathbf{q}$) which creates
(annihilates) a Tkachenko-mode excitation with wave number $\mathbf{q}$ and
energy dispersion $\epsilon_q$ \cite{matveenko2011}.  In analogy with the
derivation of Eq.~\eqref{eq:va4}, we expand $\hat{\psi}_{B}$ in
Eq.~\eqref{eq:va9} in terms of localized Wannier functions, which yields
\begin{equation}
\begin{aligned}
K_{B} + H_\mathrm{int} =
&- J\sum_{\langle i,j \rangle}\hat{b}_{i}^{\dag}\hat{b}_{j}
+ \frac{U}{2}  \sum_{i} \hat{n}_i(\hat{n}_i-1) -\mu_{B} \sum_{i}\hat{n}_i  +\\
&+
g_{AB} \sqrt{\frac{n_A}{S}} \sum_{\mathbf{q},ij}
\left[
\Lambda^{ij}_{\mathbf{q}} \hat{a}_{\mathbf{q}} +
\bar{\Lambda}^{ij}_{\mathbf{q}} \hat{a}^{\dagger}_{\mathbf{q}}
\right]
\hat{b}^{\dag}_i\hat{b}_j,
\end{aligned}
\label{eq:va10}
\end{equation}
where
\begin{equation}
\begin{aligned}
\Lambda^{ij}_{\mathbf{q}} = \int d^2r
\left[ \varphi_{A} ^{\ast}(\mathbf{r})u_{\mathbf{q}}(\mathbf{r}) -
\varphi_{A}(\mathbf{r})v^{\ast}_{\mathbf{q}}(\mathbf{r})\right]
\varphi_{B}^{\ast}(r_{i})\varphi_{B}(r_{j}),
\\
\bar{\Lambda}^{ij}_{\mathbf{q}} = \int d^2r
\left[\varphi_{A}(\mathbf{r}) u^{\ast}_{\mathbf{q}}(\mathbf{r}) -
\varphi_{A}^{\ast}(\mathbf{r}) v_{\mathbf{q}}(\mathbf{r})\right]
\varphi_{B}^{\ast}(r_{i})\varphi_{B}(r_{j}).
\end{aligned}
\label{eq:Lambdaij}
\end{equation}
$\Lambda^{ij}_{\mathbf{q}}$ is suppressed for large separations between sites
$i$ and $j$, and it can be approximately set to zero for $\left| \mathbf{R}_i -
\mathbf{R}_j\right|$ larger than $d$ (nearest-neighbor approach) -- \emph{cf.}
Appendix~\ref{AppendixMP}.  Here, however, we only consider the case of $i = j$,
that is, we also neglect the nearest-neighbor pairs $(i, j)$ in the sum in the
last term of Eq.~\eqref{eq:va10} (thus neglecting induced-tunneling terms).

The next step is to apply a unitary transformation in the total Hamiltonian
given by Eq.~\eqref{eq:va2}, to cancel the interaction term between impurities
and lattice modes. This will be incorporated in the coefficients of the impurity
BH Hamiltonian, providing an effective polaronic model \cite{demler,demler2}. We
consider
\begin{equation}
\tilde{H} = e^{-\mathcal{U}}He^{\mathcal{U}} =
H+[\mathcal{U},H]+\frac{1}{2!}[\mathcal{U},[\mathcal{U},H]]+\dots,
\end{equation}
with
\begin{equation}
\label{eq:va11}
\mathcal{U} =
\frac{1}{\sqrt{S}}\; \sum_{\mathbf{q},j} \,
\frac{1}{\epsilon_{q}}
\left[\;
\alpha_{\mathbf{q},j} ^{*}\;\hat{a}^{\dagger}_{\mathbf{q}} -
\;\alpha_{\mathbf{q},j}\; \hat{a}_{\mathbf{q}}
\right] \hat{n}_j.
\end{equation}
The transformed Hamiltonian depends on how the
impurity and lattice-modes operators are modified. For the real-space
impurity operator, we have
\begin{equation}
\label{eq:va12}
e^{-\mathcal{U}}\hat{b}_{i}e^{\mathcal{U}} =
\hat{b}_{i}\hat{X}_{i},
\end{equation}
with $\hat{X}_{i} = e^{\hat{Y}_{i}}$ and $ \hat{Y}_{i} = ({1}/{\sqrt{S}})
\sum_{\mathbf{q}} \left(\;\alpha_{\mathbf{q},i}\;\hat{a}_{\mathbf{q}} -
\;\alpha^{\ast}_{\mathbf{q},i}\;\hat{a}_{\mathbf{q}}^{\dagger}\right)$, while the
momentum-space lattice-mode operator transforms as
\begin{equation}
\label{eq:va13}
e^{-\mathcal{U}} \hat{a}_{\mathbf{q}} e^{\mathcal{U}} =
\hat{a}_{\mathbf{q}} -
\frac{1}{\sqrt{S}} \sum_{i} e^{\imath\mathbf{q}\cdot\mathbf{R}_{i}} \;
\alpha^{\ast}_{\mathbf{q},i} \; \hat{n}_{i}.
\end{equation}
By replacing the fields in Eq.~\eqref{eq:va2} with the transformed ones from
Eqs.~\eqref{eq:va12} and \eqref{eq:va13}, and choosing
$\alpha_{\mathbf{q},i}=({g_{AB}\sqrt{n_A}}/{\epsilon_{q}})
\Lambda_\mathbf{q}^{ii}$ to exactly cancel the impurity-lattice-modes
interaction, we obtain
\begin{eqnarray}
\label{eq:va14}
\nonumber
\tilde{K}^\mathrm{eff}_{B} =&&
-\tilde{J}\sum_{\langle i,j \rangle}\hat{b}_{i}^{\dagger}\hat{b}_{j} +
\frac{\tilde{U}}{2}\sum_{i}\hat{n}_i(\hat{n}_i-1) -
\tilde{\mu}_B \sum_{i}\hat{n}_i \\
&&-\sum_{\langle i, j \rangle} \frac{V_{i,j}}{2} \hat{n}_i\hat{n}_{j}.
\end{eqnarray}

Above, we neglected retardation effects, assuming that the lattice excitations
instantaneously follow the motion of a heavy B impurity. Also, to obtain
Eq.~(\ref{eq:va14}), we traced out the lattice degrees of freedom from the
transformed Hamiltonian $\tilde{K}^\mathrm{eff}_{B} = \langle
\tilde{K}^\mathrm{BOG}_{A}+\tilde{K}_{B} + \tilde{H}_\mathrm{int}
\rangle_\mathrm{ph}$, with $|\mathrm{ph}\rangle = \prod_{\mathbf{q}}
|N_{\mathbf{q}}\rangle$ and $|N_{\mathbf{q}}\rangle$ being a number state of the
lattice modes \cite{mahan}.
The effective BH parameters for the trapped species B in the
presence of the lattice modes read
\begin{align}
\nonumber
\tilde{J} &= J \langle \hat{X}^{\dag}_{i}\hat{X}_{j} \rangle_\mathrm{ph} = \\
\label{eq:va15}
&= J \; \exp\left[ -\frac{g^{2}_{AB}n_{A}}{2S}
\sum_{\mathbf{q}}
\frac{|\Lambda_\mathbf{q}^{ii}(1-e^{-\imath \mathbf{q} \cdot (\mathbf{R}_j-\mathbf{R}_i)})|^{2}}{ \epsilon_{\text{q}}^{2} }
\right], \\
\tilde{U} &=
 U - \frac{2 n_A g_{AB}^2}{ S} \sum_{\mathbf{q}}  \frac{|\Lambda_\mathbf{q}^{ii}|^2}{\epsilon_{q}}, \\
\tilde{\mu}_B &=
\mu_B +\frac{ n_A g_{AB}^2}{ S} \sum_{\mathbf{q}}  \frac{|\Lambda_\mathbf{q}^{ii}|^2}{\epsilon_{q}}.
\end{align}
Besides changes of the Bose-Hubbard-model parameters ($J$, $U$, and $\mu_B$),
the lattice dynamics also induces an attractive long-range potential between the
impurities, which is mediated by the lattice modes. This corresponds to the last
term in Eq.~\eqref{eq:va14}, where
\begin{eqnarray}
\label{eq:va17}
V_{i,j}
 =
\frac{2 n_A g_{AB}^2}{S}
\sum_{\mathbf{q}}
\frac{
\left(\Lambda^{ii}_\mathbf{q}\right)^\ast \;
\Lambda^{jj}_\mathbf{q}
}{\epsilon_{q}}.
\end{eqnarray}
Based on the expressions for the Bogoliubov coefficients $u_{\mathbf{q}}$ and
$v_{\mathbf{q}}$ at small $q$, we can estimate the strength $V_{i,j}$ of the
isotropic long-range attraction between atoms on neighboring sites $i$ and $j$,
at a lattice distance $d$.  By using the quadratic Tkachenko-mode dispersion $
\epsilon_{\text{q}}$ \cite{pereira}, we find  (see Appendix \ref{AppendixMP} for
a detailed derivation)
\begin{equation}
V_{i,j} \propto \frac{e^{-{d^2}/{2\ell_0^2}}}{d^2}.
\end{equation}
The strong suppression of $V_{i,j}$ at large distance justifies considering a
truncated model, in which $V_{i,j}={V}$ when $i$ and $j$ are nearest neighbors,
and $V_{i,j}=0$ otherwise.  The parameter ${V}>0$ quantifies the strength of the
nearest-neighbor attraction, and the departure from the ordinary BH model.

\section{Extended Bose-Hubbard phase diagram}
\label{sect:EBH}

In this section, we characterize the phase diagram of the resulting EBH model
described in Sec.~\ref{sect:dynamics} by Eq.~\eqref{eq:va14}.  We first treat
the case of vanishing hopping parameter $\tilde{J}$ (atomic limit), and we
assume $z {V}<\tilde{U}$ to guarantee the system stability. For larger values
of the nearest-neighbor attraction ${V}$, it is energetically favorable to add
an infinite number of particles in the system.

\begin{figure}[hbt]
\centering
\includegraphics[width=0.9\linewidth]{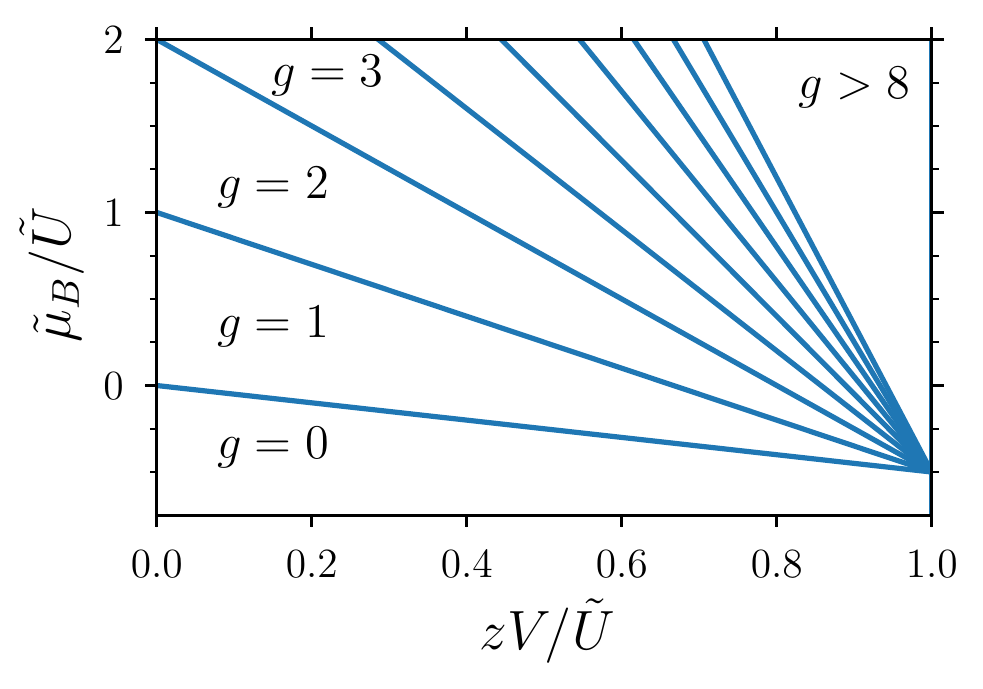}
\caption{Phase diagram of the EBH model (\emph{cf.} Eq.~\eqref{eq:va14}) in the
atomic limit ($\tilde{J}=0$), formed by Mott-insulator regions with integer
filling factor $g$.  Phase boundaries are obtained through
Eq.~\eqref{eq:muggplus1} (regions with filling $g>8$ are merged together, for
clarity). The system is unstable for $z {V}/\tilde{U}>1$.
\label{fig:atomic_limit}}
\end{figure}

The simplest ground state ansatz is a homogeneous state with $g$
particles per site:
\begin{equation}
\label{eq:Psi_g}
\left|\Psi_g \right\rangle =
\frac{1}{\sqrt{g!}} \sum_{i} \big(\hat{b}_i^\dagger\big)^g \left|0\right\rangle,
\end{equation}
where $i$ runs over all lattice sites.
By minimizing $\left\langle \Psi_g \right| K_B^\mathrm{eff} \left| \Psi_g
\right\rangle$ with respect to $g$, we obtain the ground-state filling factor.
The boundary between the regions with $g$ and $g+1$ atoms per site is given by
\begin{equation}
\tilde{\mu}_B^{(g,g+1)} = g \tilde{U} - \frac{2g+1}{2} z {V},
\label{eq:muggplus1}
\end{equation}
and the resulting phase diagram is represented in Fig.~\ref{fig:atomic_limit}.
The Mott-insulator regions of the ordinary BH phase diagram (${V}=0$) remain present for
all values of ${V}$, but their size and position are modified. The
size of each Mott region decreases as $V$ increases, with the extension of the
$g$-th region along the $\mu_B$ axis being equal to $\tilde{U}-zV$.  Moreover,
the phase boundaries have a negative slope as a function of $V$. This shift
follows from the fact that the nearest-neighbor attraction acts as an additional
chemical potential, favoring the addition of more atoms in the lattice.

\begin{figure}[hbt]
\centering
\includegraphics[width=0.5\linewidth]{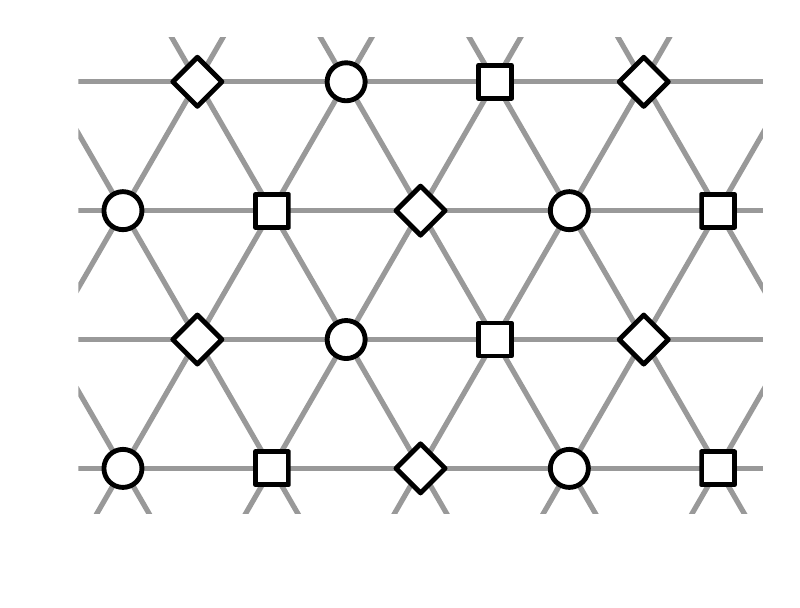}
\caption{Partition of a triangular lattice into three sub-lattices, represented
with different symbols.
\label{fig:sublattices}}
\end{figure}

The ansatz $|\Psi_g\rangle$ cannot reproduce inhomogeneous states, like the
charge-density-wave identified for repulsive nearest-neighbor interactions.
More general ans\"{a}tze may be used, that include inhomogeneous state. For
instance one can consider the state $\left| \Psi_{g_1, g_2, g_3} \right\rangle$,
with integer filling factors $g_1, g_2,$ and $g_3$ on the three sub-lattices
represented in Fig.~\ref{fig:sublattices}. The numerical minimization of the
variational energy $\left\langle\Psi_{g_1, g_2, g_3}\right| K_B^\mathrm{eff}
\left|\Psi_{g_1, g_2, g_3}\right\rangle$ leads to $g_1 = g_2 = g_3$, implying
that the ground state falls in the class of homogeneous states -- \emph{cf.}
Eqs.~\eqref{eq:Psi_g} and \eqref{eq:muggplus1}.  Note that, for repulsive
nearest-neighbor interactions (not treated in this work), the tripartite ansatz
$\left|\Psi_{g_1, g_2, g_3}\right\rangle$ would also produce ground states with
fractional filling factors (e.g., with $g_1 = 1$ and $g_2 = g_3 = 0$).

For $\tilde{J} = 0$, the EBH Hamiltonian is diagonal in the basis of states with
fixed filling factors on all sites, so that finding the ground state for a
$L\times L$ finite lattice corresponds to the optimization of a function of
$L^2$ variables (that is, the local occupation numbers).
We address this multidimensional optimization problem through the
simulated-annealing algorithm \cite{Kirkpatrick1983} for extended systems
($L\geq12$).  The numerical results confirm that the ground state is
homogeneous, with filling factor determined by Eq.~\eqref{eq:muggplus1}.

\begin{figure*}[bt]
\centering
\includegraphics[width=0.98\linewidth]{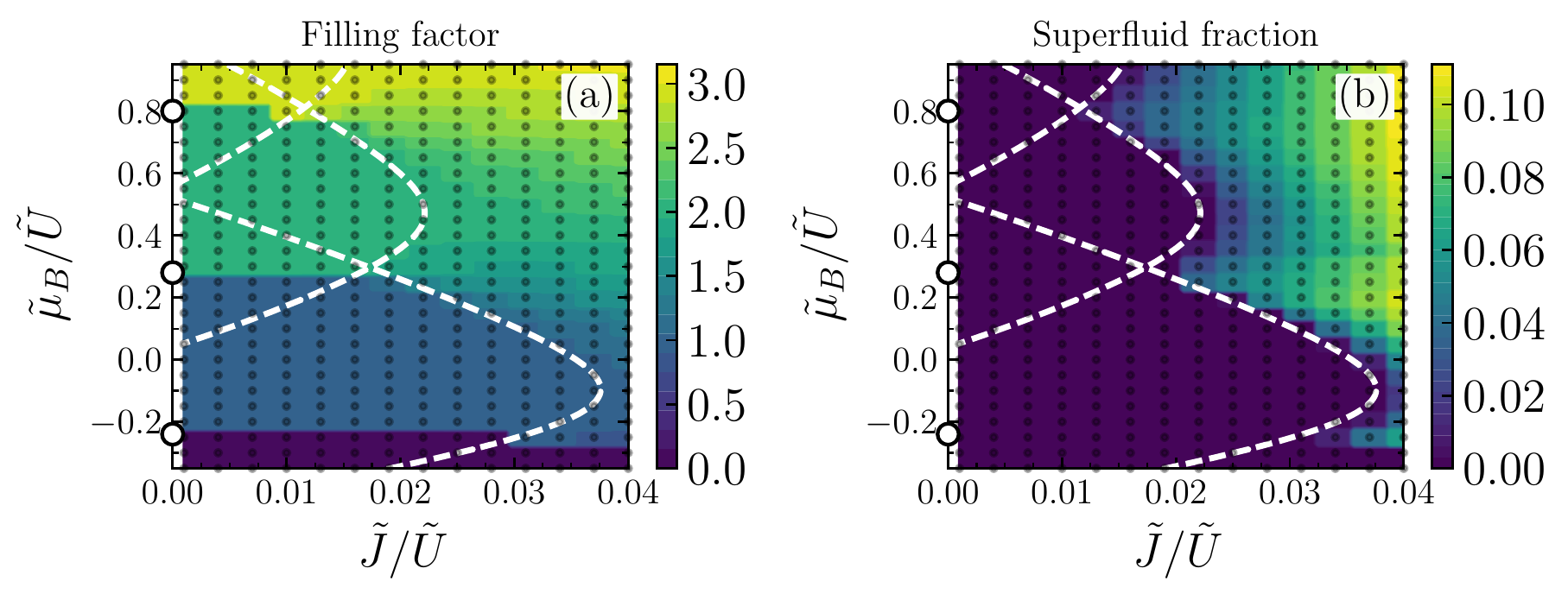}
\caption{Average filling factor $\rho$ (a) and superfluid fraction $\rho_s/\rho$
(b) for the Hamiltonian in Eq.~\eqref{eq:va14}, with nearest-neighbor attraction
${V} / \tilde{U}=0.08$.  Color code: QMC data for a lattice of $18\times18$
sites with periodic boundary conditions, at temperature $T = \tilde{U} / 20$
(simulation are performed at points marked by gray dots).  Dashed white line:
Phase boundaries of the Mott insulators with $g=1,2,3$, obtained by shifting the
${V}=0$ data from Ref.~\cite{Teichmann2010EPL} -- \emph{cf.}
Eq.~\eqref{eq:shifted_boundaries}.
\label{fig:qmc}}
\end{figure*}

The two variational ans\"{a}tze and the direct-optimization method employed in
the atomic limit cannot be directly generalized to the $\tilde{J} > 0$ case,
where the Hamiltonian also includes off-diagonal terms.  To determine the phase
diagram in this region (see Fig.~\ref{fig:qmc}), we compute the average filling
factor (i.e., the density $\rho$) and superfluid fraction $\rho_s / \rho$
through QMC simulations of large systems at low temperature, effectively probing
the ground state.  At small $\tilde{J} / \tilde{U}$, we observe a direct
transition between MI states with different filling factors (\emph{cf.}
Fig.~\ref{fig:qmc}(a)), with the phase boundaries given approximately by
Eq.~\eqref{eq:muggplus1} .

For larger values of the hopping coefficient, the transition towards the
superfluid phase is signaled by the superfluid fraction acquiring a finite
value.  The position of this transition, for $V>0$, is well captured by a
shifted version of the $V=0$ phase boundary.  If we denote the critical hopping
by $\tilde{J} / \tilde{U} = f(g, \tilde{\mu}_B / \tilde{U}, {V}/\tilde{U})$,
then Fig.~\ref{fig:qmc}(b) suggests that the simple relation
\begin{equation}
f\left(g, \frac{\tilde{\mu}_B}{\tilde{U}}, \frac{V}{\tilde{U}} \right) =
f\left(g, \frac{\tilde{\mu}_B - g z V}{\tilde{U}}, 0 \right)
\label{eq:shifted_boundaries}
\end{equation}
holds close to the tips of MI lobes, that is, for large-enough $\tilde{J} /
\tilde{U}$. The right-hand side in Eq.~\eqref{eq:shifted_boundaries} can be
obtained from the process-chain results \cite{Teichmann2010EPL}.

\section{Experimental considerations}
\label{sect:experiments}

We now discuss the experimental feasibility of our proposal, starting from the
criteria on the choice of the mixture.  The first required condition is that
$m_B \gg m_A$, since the lattice potential felt by the impurities (species B)
must be deep enough that they can be trapped by the vortex cores. This condition
is also fulfilled with $g_{AB}$ sufficiently large.

For species A, we consider bosonic atoms for which the formation of highly
ordered vortex lattices has been observed experimentally \cite{fetter1}, namely
the alkalis Na, Li, and Rb.  We limit our discussion to the commonly explored
mixture of $^{23}$Na and $^{87}$Rb, to realize the vortex lattice and heavy
impurities, respectively.  The BEC lifetime is limited by three-body losses.
Requiring that it exceeds a few tens of seconds, we choose $n_{A} \approx
(10^{20} /\mathrm{m}^3 ) \times \ell_z$, with an effective BEC thickness $\ell_z$ $
(\omega_{z}\sim 2\pi\times 5$ kHz).  Assuming that the bare (non-resonant)
scattering length interaction is $a_A = 60 a_0$, where $a_0$ is the Bohr radius,
the chemical potential reads $\mu_A = 1.4$ kHz, ($\sim 70$ nK).  To form a
vortex lattice that is sufficiently large, homogeneous and stable, we consider
similar values for the rotation and the interaction energies ${n_{A} g_A} / {2
\hbar \Omega} \sim 1$.  At this critical array vorticity $\Omega$, the residual
radial trap vanishes in Eq.~\eqref{eq1} (the harmonic oscillator length of
$V_{\textrm{ext}}(\mathrm{r})$ coincides with the magnetic length $\ell =
\sqrt{\hbar / (m_A \Omega)}$) and one has tightly packed triangular
vortex-lattice geometry, with lattice parameter $d = 2 \ell \sim 1.6$ $\mu$m and
sites of size $\xi \sim 0.4$ $\mu$.

For species B, the recoil energy $E_r$ of an impurity localized in the vortex
core has to be much smaller than the potential barrier of magnitude $V_0$.
Taking into account all considerations elaborated above, we find a reasonable
value $V_0 / E_r \sim 12$ for $a_{AB} = 240 a_0$. It is important to stress that
with our chosen parameters, the vortex-lattice lifetime can be of the order of
several seconds, while the characteristic time associated with the tunnelling
$J$ in Eq.~(\ref{eq:va6}) is approximately $0.2$ s. In addition, we assume that
the occupation in the sites are low enough that the states of the impurities are
accurately described using the lowest-band Wannier functions.  Reminding that
the density of sites $n_V$ is constraint by $n_A$ in the high filling-factor
regime, $\nu = N_A/N_V \gg 1$, the impurities have a negligible influence on the
lattice bosons, due to the diluteness of species B ($n_B\sim n_V$).

We can then estimate the characteristic energies for the Bose-Hubbard
coefficients of species B in Eq.~\eqref{eq:va4}. For the hopping parameter we
have $J \sim 0.004 \;\mu_A$. In the two-species vortex-lattice setup, the ratio
$U / J$ can be tuned by changing the scattering length $a_B$, as shown by
Eq.~\eqref{eq:va7}. To access the region of the phase diagram that corresponds
to the MI-SF transition, assuming the unitary occupation of the lattice sites,
we consider $a_B^\mathrm{crit} \approx 277 a_0$, which gives ${U}/{J} \sim26.6$.

The extended Bose-Hubbard (EBH) in Eq.~\eqref{eq:va14}, includes an attractive
long-range potential with magnitude $V \sim 0.005 \;\mu_A$ (\emph{cf.}
Eq.~\eqref{Ap4} in Appendix \ref{AppendixMP}).  Importantly, we have ${V} /
\tilde{U}\sim 0.05$, which obeys the stability condition for the ground-state
solution of the EBH, $z {V} / \tilde{U} < 1$.

Using a $^7$Li-$^{133}$Cs mixture would require a magnetic field on the order of
$850$ G to render the scattering length of $^7$Li positive and sufficiently
large. An advantage of this mixture would be the high value of the ratio $V_0 /
E_r$ which could be obtained without significantly increasing $a_{AB}$.  Working
with a more massive impurity than $^{87}$Rb would allow one to realize the MI-SF
transition using smaller values for the scattering lengths. The disadvantage,
however, would be the higher three-body loss rate estimated for $^7$Li
\cite{servaas}, reducing lifetime for the vortex lattice.

\section{Conclusions}
\label{sect:conclusions}

In the present work, we propose a setup to realize a BH model with a mixture of
ultracold atomic gases in the presence of an effective rotation (namely, an artificial gauge-field).  Similarly to the typical optical-lattice setups, the
tunability of the physical parameters of the system allows to explore a wide
range of regimes for the effective BH model.  The vortex-lattice
Tkachenko modes, in particular, modifies the BH parameters and introduces an
additional long-range attraction.  The EBH model with long-range repulsion
(stemming from strong dipolar interatomic interactions) has been studied in
detail, and it led to the prediction of additional phases which are not present
in the ordinary BH case, like the density-wave and supersolid phases in two
dimensions \cite{lewenstein2, yue}.  For the attractive EBH model in
Eq.~\eqref{eq:va14}, we compute the phase diagram through the quantum Monte
Carlo technique. The Mott-insulator regions of the BH case remain in the
extended model, but their size and position are modified by the long-range
attraction.

A recent publication \cite{jaksch} treats a similar topic, that is, the phase
diagram of impurities in a vortex lattice. However, Ref.~\cite{jaksch} concerns
the vortex-lattice deformations caused by the strong interaction with the
impurity, and how the higher occupation of the sites affects the BH parameters.
In this work, in contrast, we studied the weak-coupling limit, focussing on the
interaction of the impurity with the vortex-lattice degrees of freedom. After
establishing the stability conditions for a vortex lattice in the presence of
multiple impurities, we explored the effects of the lattice dynamics on the
confined species. Treating the dilute system by means of an effective polaronic
Hamiltonian, we showed how it allows to go beyond the present studies with atoms
trapped in static optical lattices.

Besides being part of an unusual BH class, our proposed model is also an
interesting experimental proposal in the context of Bose-Einstein condensate
mixtures, since it requires the application of the most recent and successful
techniques in the ultracold atoms field: The advances in cooling mechanism to
produce binary condensates \cite{odelin}, the Feshbach-resonance technique to
control the interaction parameters, and artificial gauge-fields to selectively
nucleate vortex lattices in one of the atomic species \cite{spielman}.  The
possible quantum phases can be characterized through the spatial noise
correlations in the absorption image of the free expanded atomic cloud
\cite{lukin}. Similar techniques can be used to observe signatures of more
exotic states, such as spin liquids \cite{poland}.

Fermions could also be considered in this same framework. Using two hyperfine
states, interesting effects are expected to arise in a Mott-insulator phase of
the pseudo-spin fermions, with the interplay between the triangular-lattice
geometry and spin ordering frustration \cite{Lewenstein2012, esslinger2}. In the
optical lattice, however, the study of quantum spin models and of
strongly-correlated magnetic phases, as the spin-liquid phase, has been limited
by the high temperature of the fermions in the lattice, which is still higher
than that required to observe exchange-driven spin ordering. In our case, in
contrast, the BEC vortex-lattice background can act like a reservoir, with the
cooling of the trapped fermions coming from the creation of excitations in this
reservoir. This favorable scenario was proposed in the context of dissipative
Hubbard models \cite{zoller4}, where the strong control over many parameters, as
interactions between atoms, allows one to manipulate the system-reservoir
coupling.

\begin{acknowledgments}
We thank Marco Di Liberto for insightful comments, and Martin Holthaus for
providing the data from Ref.~\cite{Teichmann2010EPL}.  This work is supported by
CNPq and FAPESP.
\end{acknowledgments}

\onecolumngrid

\appendix

\section{Mediated potential}
\label{AppendixMP}

Based on the Bogoliubov transformation of the vortex-lattice
\cite{matveenko2011}, we can determine the profile of the long-range effective
potential given by Eq.~(\ref{eq:va17}) in the main text. In particular, its
asymptotic behavior can be estimated analytically with the low-energy
(long-wavelength) Tkachenko modes contribution, as will be shown.

As studied before \cite{pereira}, for $q\ll  \ell^{-1}$, with the magnetic
length $\ell$ related to the inter-vortex separation through $d=2\ell$, the
gapless Tkachenko modes have dispersion relation $\epsilon_{q} \approx
\hbar^2q^2/2M$, with $M= \frac{1}{2\kappa\sqrt{\eta}}\frac{\hbar\Omega}{n_A
g_A}m_A$ and the lattice constant $\kappa = 1.1592$ and  $\eta = 0.8219$. In the
low-energy limit, we have $u_{\mathbf q}(\mathbf r)\approx \varphi_A(\mathbf
r)\; c_{1\mathbf q}\,e^{\imath \mathbf q\cdot \mathbf r}$ and $v_{\mathbf q}(\mathbf
r)\approx \varphi_A(\mathbf r)\; c_{2 \mathbf q}\,e^{-\imath \mathbf q\cdot \mathbf
r}$. The small value of the momentum allows us to expand $(c_{1q}-c_{2q}) \approx
\frac{1}{\sqrt{2}}\,\eta^{1/4}\,(q\, \ell)$.

According to the Extended Bose-Hubbard Hamiltonian (Eq.~\eqref{eq:va14}), we can
determine the profile of the mediated potential from Eq.~\eqref{eq:Lambdaij},
assuming again the gaussian function for $\varphi_{B}(r)\,(=
|B_{0}|\,e^{-r^2/2\ell_{0}^{2}}\,$)
\begin{eqnarray} \nonumber \label{Ap4}
V_{i,j} = V(d)&\sim& \frac{2 n_A g_{AB}^2}{S}   \sum_{\mathbf{q}} \frac{\left[\frac{1}{\sqrt{2}}\,\eta^{1/4}\,(q\, l)\right]^2}{\epsilon_{q}} \bigg[\int d^2r |B_{0}|^2 \,e^{-|\mathbf{r}-\mathbf{d}|^2/2\ell_{0}^{2}}\;e^{-r^2/2\ell_{0}^{2}}\bigg]^2 \\ \nonumber
&=& \frac{g_{AB}^2}{g_A}\frac{1}{\kappa} |B_{0}|^4  e^{-d^2/\ell_0^2}\int_{0}^{2\pi/\ell} q dq \bigg[\int r dr e^{-r^2/\ell_0^2}\int_{0}^{2\pi} d\theta e^{r d  \cos\theta/\ell_0^2}\bigg]^2 \\
&=&  \frac{g_{AB}^2}{g_A}\frac{4\pi}{\kappa} \frac{1}{d^2}e^{-d^2/2\ell_{0}^{2}}.
\end{eqnarray}
This result justifies the restriction of the potential range to
nearest-neighboring sites.

\section{Details of the quantum Monte Carlo calculations}
\label{AppendixQMC}

The quantum Monte Carlo simulations in this work make use of the worm algorithm
\cite{Prokofev1998PLA}, as implemented in the ALPS libraries
\cite{Albuquerque2007JMMM, Bauer2011JStatMech}.  For the bosonic models in
Eqs.~(\ref{eq:va4}) and (\ref{eq:va14}), this represents an unbiased method to
compute observables as the density or energy, at finite temperature $T$ and for
a finite number of sites $L^2$.

\begin{figure}[hbt] \centering
\includegraphics[width=0.55\linewidth]{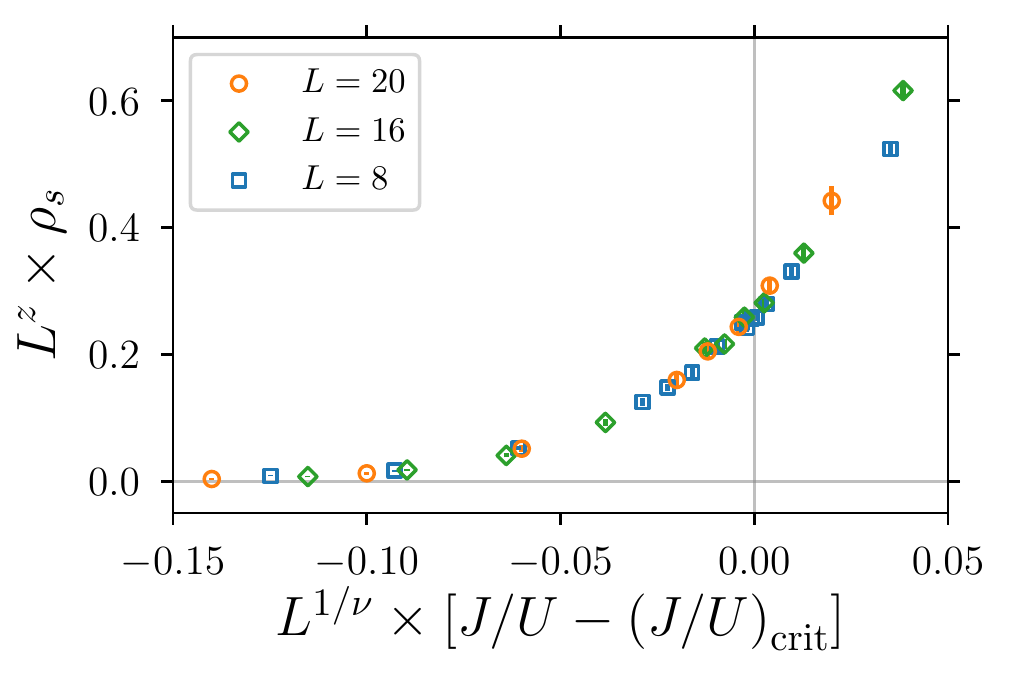}
\caption{
Data collapse of the superfluid stiffness $\rho_s$: Data for different linear
sizes $L$ (see legend) are rescaled as in Eq.~(\ref{eq:fss}), with
$z=2$ and $\nu=1/2$ being the dynamical and correlation-length critical
exponents.  With the product $T\times L^z$ kept equal to 0.5,
different lines collapse onto a single universal curve.  Data are shown for
chemical potential $\mu_B/U=0.645$, with $(J/U)_\mathrm{crit}=0.02595$.
\label{fig:collapse}}
\end{figure}

The phase boundary showed in Fig.~1 is obtained through the finite-size scaling
technique \cite{fisher}, by finding the critical hopping parameter
$J_\mathrm{crit}$ for fixed $U$ and $\mu_B$.  Away from the tip of the
Mott-insulator lobe, a \emph{generic} phase transition is expected, with
dynamical and correlation-length critical exponents equal to 2 and $1/2$,
respectively \cite{fisher, Krutitsky2016PR}. Under this assumption, and for $J$
close to $J_\mathrm{crit}$, the superfluid stiffness $\rho_s$ satisfies
\begin{equation}
L^2 \rho_s = F\left(\frac{J_\mathrm{crit}-J}{J_\mathrm{crit}} L^2, \frac{1}{L^2 T} \right),
\label{eq:fss}
\end{equation}
where $F$ is a universal function. Thus $J_\mathrm{crit}$ can be obtained by
plotting $L^2 \rho_s$ as a function of $J$ for several linear sizes $L$, at
fixed $T \times L^2$, and by extracting the common intersection point of these
lines (see for instance Ref.~\cite{Smakov2005PRL}).  When rescaled as in
Eq.~(\ref{eq:fss}), superfluid-stiffness lines corresponding to different
values of $L$ collapse onto a single curve, confirming the correct estimate of
the critical value $J_\mathrm{crit}$ -- see the example in
Fig.~\ref{fig:collapse}.

\end{document}